\documentclass[aps,floats,amssymb,twocolumn,prb,superscriptaddress,showpacs,floatfix]{revtex4}
\usepackage{calc}
\usepackage{psfrag}
\usepackage{graphicx}

\begin{document}
\title{Transition from two-component 332 Halperin state to  one-component Jain state at filling factor $\nu=2/5$ }
\author{M. V. Milovanovi\'c}
\affiliation{Scientific Computing Laboratory, Institute of Physics, University of Belgrade, P.~O.~Box 68, 11 000 Belgrade, Serbia}
\author{Z. Papi\'{c}}
\affiliation{Scientific Computing Laboratory, Institute of Physics, University of Belgrade, P.~O.~Box 68, 11 000 Belgrade, Serbia}
\affiliation{Laboratoire de Physique des Solides, Univ.~Paris-Sud, CNRS, UMR 8502, F-91405 Orsay Cedex, France}

\begin{abstract}
We study the transition induced by tunneling from the two-component
332 Halperin's state to the one-component Jain's state at the
filling factor $\nu=2/5$. In exact diagonalizations of small systems
two possibilities for the transition are found: (a) avoided level
crossing, and (b) level crossing i.e. first-order transition in the
case of Coulomb interaction and short range interaction,
respectively. An effective bosonic model with $p$-wave pairing for
the transition is proposed. The relevance of the Gaffnian state for
the transition is discussed as well as possible consequences of our
model on the effective description of the Jain's state.
\end{abstract}


\maketitle \vskip2pc

\section{Introduction} \label{sec_introduction}

Topological phases of matter\cite{Wen} find their concrete realizations
in quantum Hall physics within the systems of 2D electrons in high magnetic fields.
They are characterized by a gap to all excitations and by degeneracy of the ground state
on higher genus surfaces. By changing a parameter of the electronic system, we may
induce a quantum phase transition from one topological phase to the other. Due
to their nature and discrete characterization, we expect that
the system gap closes at the transition between topological phases that differ in
the topological invariants i.e. the numbers that characterize them.

The fractional quantum Hall (FQH) states can be characterized by the
filling factor $\nu$ i.e. particular ratio between the density of
electrons and the strength of the magnetic field at which they
appear. In spin-polarized systems, a successful explanation of
various FQH states at different filling factors is given by Jain's
states.\cite{JJain} On the other hand, for the same filling factors
we may have states with two or more different species i.e. the
Halperin states. \cite{halperin} Usually for a given filling factor
described by the Jain's state, the corresponding Halperin state has
the same vacuum degeneracy but some other characteristic numbers may
differ. By applying tunneling to a two-component Halperin state we
may transform this state into the one-component (spin-polarized)
Jain's state. Tunneling as a perturbation that drives the transition
from the two-component to a one-component FQH system was studied
previously by analytical\cite{wen_prl,read_green} and
numerical\cite{mcdonald_haldane} means.

In this paper we study the transition from Halperin's two-component
332 state to the one-component Jain state at the filling factor $\nu
= 2/5$ via tunneling. The interest is threefold: we would like to
find out (a) about the nature of quantum phase transitions between
topological phases which are similar (332 and Jain's
state have the same ground state degeneracy\cite{haldanetorus,wzmatrix,wencsl} but different
shift\cite{hierarchy, wen_shift}), (b) we would like to find if
Gaffnian\cite{SS} can be characterized as a critical state in these
circumstances when the gap closes, and (c) we explore possible
consequences for the effective description of the Jain state due to
a better understanding of the transition. In Sec.\ref{sec_system} we
define the electronic system that we consider. Sec.\ref{sec_ed}
contains the results of the exact diagonalization studies of the
transition. In Sec.\ref{sec_bosonic} we introduce an effective
bosonic model of the system and  the transition induced by
tunneling.  Sec.\ref{sec_conclusions} is devoted to conclusions.

\section{The system under consideration}\label{sec_system}

We consider the quantum Hall bilayer in the presence of the vector potential
$\mathbf{A}$ that describes a strong magnetic field, $B\hat z =
\nabla\times \mathbf{A}$, perpendicular to both layers.
In the rotationally symmetric gauge, the LLL
eigenstates of an electron with the coordinate $z = x + i y$ in the plane and
localized in the layer $\sigma \in \{ \uparrow, \downarrow \}$ are given by
\begin{equation}
z^{m} \exp\{- |z|^{2}/4l_B^2\}\eta_\sigma,\;\;\;m = 0,\ldots,N_{\phi}-1.
\label{eis}
\end{equation}
where $\eta_\sigma$ is the usual spinor wave function and the unit of length is given by the
magnetic length, $l_{B} = \sqrt{\hbar c/eB}$. The number of flux quanta, $N_\phi$,
denotes the number of available states in the lowest Landau level (LLL). In the thermodynamic
limit, the ratio of the number of electrons $N_e$ and the number of flux quanta $N_\phi$
defines the filling factor $\nu=N_e/N_\phi$ and we focus on the particular case $\nu=2/5$.

The many-body interacting system of electrons is defined by the following Lagrangian
density in the second quantized formulation:
\begin{eqnarray}
\nonumber {\cal L}&=& \sum_{\sigma}\Big\{\Psi^{\dagger}_{\sigma}\partial_{\tau}
\Psi_{\sigma} - \Psi^{\dagger}_{\sigma}\frac{(\partial_{\mathbf{r}} + e
\mathbf{A})^{2}}{2 m}
\Psi_{\sigma}  - \Psi_\sigma^\dagger \frac{\Delta_{SAS}}{2}  \Psi_{-\sigma}\\
&&+ \frac{1}{2}\int\; d\mathbf{r}' \rho_{\sigma}(\mathbf{r})
V_{c}^{\rm intra}(\mathbf{r}-\mathbf{r}') \rho_{\sigma}(\mathbf{r}')\nonumber \\
&& + \frac{1}{2} \int\; d\mathbf{r}' \rho_{\sigma}(\mathbf{r})
V_{c}^{\rm inter}(\mathbf{r}-\mathbf{r}')
\rho_{-\sigma}(\mathbf{r}')\Big\} , \label{def332}
\end{eqnarray}
where $\Psi_{\sigma}$ is the electron field which carries the pseudospin (layer) index
and $\Delta_{SAS}$ denotes the tunneling term.
The interaction is defined by
\begin{equation}\label{intra}
V_{c}^{\rm intra} (r) = \frac{e^{2}}{\epsilon r}
\end{equation}
and in general $V_{c}^{\rm inter}$ is different. When we model a quantum
Hall bilayer,
\begin{equation}\label{inter}
V_{c}^{\rm inter} (r) = \frac{e^{2}}{\epsilon \sqrt{r^{2} + d^{2}}},
\end{equation}
$d$ has the meaning of the distance between two layers of
two-dimensional gases and it is of the order of $l_B$.
In the Lagrangian density (\ref{def332}) and the remainder of this paper
we set $\hbar = c = l_B = 1$. Significant insight into the physics described by the Lagrangian (\ref{def332})
can be obtained using first-quantized trial wave functions for its ground states. \cite{laughlin}
In the remainder of this section we list several candidate wave functions that are expected to describe the
ground state of (\ref{def332}) in different limits of  $\Delta_{SAS}$ and $d$.
Trial wave functions in the LLL are analytic in $z$ variables and we will omit the omnipresent
 Gaussian factor for each electron as the one in (\ref{eis}).

In the small tunneling regime, the FQH system at $\nu=2/5$ is two-component,
described by the 332 Halperin state for two distinguishable
species of electrons,
$z_{i \sigma}; \sigma = \uparrow,\downarrow; i =
1,\ldots,N_{e}/2$
\begin{equation}\label{332_state}
\Psi_{332} = \prod_{i<j}(z_{i\uparrow} - z_{j\uparrow})^{3}
\prod_{k<l}(z_{k\downarrow} - z_{l\downarrow})^{3}
\prod_{p,q}(z_{p\uparrow} - z_{q\downarrow})^{2}.
\end{equation}
Due to the fact that the correlation exponents between electrons
of the same layer are bigger than those between electrons of the
opposite layers, we expect the wave function (\ref{332_state}) to
be more appropriate for non-zero $d$ e.g. in the range $d \sim l_B$.
However, as it possesses the necessary symmetry properties, \cite{hierarchy}
it can be a candidate also for $d=0$. The properties of the wave function
(\ref{332_state}) were numerically verified in Ref. \onlinecite{ymg}.

As the tunneling strength $\Delta_{SAS}$ is increased, the electrons find it
energetically favorable to be in the superposition of two layers, $\uparrow + \downarrow$,
and the system loses its two-component character. The effective single-component state
is characterized by full polarization in the $x$-direction. At $\nu=2/5$ in the LLL, a compelling
candidate for the polarized state is Jain's composite fermion (CF) state \cite{JJain}:
\begin{equation}
\Psi_{\rm Jain}= {\mathcal P}_{LLL}  \Big\{ \prod_{i<j} (z_{i} -
z_{j})^{2} \,\cdot\, \chi_{2} (\{z\}) \Big\}\, , \label{jain}
\end{equation}
where ${\mathcal P}_{LLL}$ is a projector to the LLL and $\chi_{2}$
represents the Slater determinant of two filled pseudo-Landau
levels of CFs. \cite{JJain} Note that a single index suffices to label the electron
coordinates as the pseudospin index is implicitly assumed to be $\uparrow+\downarrow$.

Recent work \cite{SS} has introduced an alternative candidate for the polarized state at the
filling factor $\nu=2/5$, the so-called Gaffnian state:
\begin{equation}
\Psi_{\rm Gaff}={\cal A}\Big\{\Psi_{332}\;\; {\rm perm}\{\frac{1}{z_{\uparrow} -
z_{\downarrow}}\}\Big\}. \label{gaff}
\end{equation}
In the notation of Eq. (\ref{gaff}) one can think of the Gaffnian originating from the
two-component 332 state with the additional pairing represented by the permanent,
a determinant with plus signs. \cite{Yo, RGJ} The two-component state is made single-component
under the action of the antisymmetrizer $\cal{A}$ between $\uparrow$ and $\downarrow$
electron coordinates. Gaffnian (\ref{gaff}) has generated a surge of interest because in finite
size (spherical) exact diagonalization it shows high overlaps with the Coulomb ground state,
comparable to those of Jain's state, yet the topological properties of the two states are very
different. \cite{SS} Moreover, the strong evidence for Gaffnian in numerical calculations is
puzzling in view of the fact that it is a correlator of a non-unitary conformal field theory
and hence not expected to describe a stable phase. \cite{read_viscosity}
In the spherical geometry, Jain's state and the Gaffnian can only be distinguished by their
excitation spectrum \cite{tokej} or by using advanced tools such as the
entanglement spectrum. \cite{rbh}

Since the antisymmetrizer $\mathcal{A}$ can, to some extent, be mimicked by the tunneling term, \cite{zetal}
and since the Gaffnian incorportates the pairing defined by the permanent, there is an additional natural candidate
which we refer to as  the permanent state,
\begin{equation}\label{perm_state}
\Psi_{\rm perm}=\Psi_{332}\;\; {\rm perm}\{\frac{1}{z_{\uparrow} - z_{\downarrow}}\}.
\end{equation}
This state distinguishes between $\uparrow$ and $\downarrow$ electrons, hence it is expected in the limit
of intermediate tunneling $\Delta_{SAS}$ before a full $x$-polarization has been achieved.
Like the Gaffnian, the state (\ref{perm_state}) is related to a non-unitary conformal field theory \cite{rr}
and one may expect that it plays a role of the critical state in the transition region before full $x$-polarization.

In the following Section we study numerically the transitions between two-component and one-component states
at the filling factor $\nu=2/5$ via tunneling $\Delta_{SAS}$. We use the exact diagonalization in
the spherical and torus geometries to gain complete insight into topological properties of
the different competing trial states introduced here.

\section{Exact diagonalizations}\label{sec_ed}

We consider the transition from the 332 (two-component) Halperin
state to the onecomponent state at $\nu = 2/5$ via tunneling. The
onecomponent state is identified below as Jain's (Abelian) state
(\ref{jain}), though it is not at the same shift on the sphere as
the 332 state. \cite{hierarchy, wen_shift} The shift $\delta =
N_{e}/\nu-N_{\phi}$ is a topological number
\cite{Wen,read_viscosity} and defined through a relation between
$N_{e}$ and $N_{\phi}$ that is necessary for the appearance of a
particular FQH state on the sphere. For example, in case of the 332
state $\delta=3$, whereas for the states in Eqs. (\ref{jain},
\ref{gaff}, \ref{perm_state}) $\delta=4$. This mismatch is an
unfortunate feature of the spherical geometry which prevents the
direct study of the transition. However, all of the mentioned states
describe the filling $\nu=2/5$ and therefore occur in the same
Hilbert space under the periodic boundary conditions where the shift
is trivially zero. \cite{haldane_rezayi_ed, hierarchy}(By that the
phases in the torus geometry do not ``loose" the topological number
connected with the shift on the sphere, this number that reflects
the orbital spin can be characterized by the Hall viscosity of the
system. \cite{read_viscosity})  Thus in the torus geometry we can
study the transitions in a direct manner. As we mention below,
another advantage of the torus geometry is the specific ground-state
degeneracy which can be used as a fingerprint of a phase. The
physical results derived from the two geometries, however, ought to
agree for large enough systems.  Our numerical studies are
restricted to a small number of electrons because the tunneling does
not conserve particle number in each layer. Since we anticipate
incompressible states for most of the range of $\Delta_{SAS}$, small
system sizes are nonetheless expected to be relevant as usual in the
context of quantum Hall effect. \cite{hierarchy}

\subsection{Sphere}\label{sec_sphere}

In the spherical geometry, Coulomb or any interaction that depends
on the distance between two electrons is parameterized  by a
discrete series of the so-called pseudopotentials in the
LLL.\cite{hierarchy} Each pseudopotential is an eigenvalue of the
interaction strength corresponding to the state of definite relative
angular momentum ($l$) of two electrons. Therefore a series of
pseudopotentials $\{ V_{l} | l =0, 1,\ldots \}$ completely specifies
the interaction in the LLL. Model pseudopotentials define an
interaction in the LLL for which the analytic functions of some
simple fractional quantum Hall states are the densest zero energy
eigenstates. This is the case for the 332 state when $V^{\rm intra}
= \{0, V_{1}^{a},0,0,\ldots\}$ and  $V^{\rm inter} = \{V_{0},
V_{1}^{e},0,0,\ldots\}$. There is some freedom in choosing $V_0,
V_1^{a,b}$ apart from the requirement that they should all be
positive and  we set them to unity. Values of $V_0, V_1^{a,b}$
control the gap for the 332 state and thereby affect the critical
value for the tunneling $\Delta_{SAS}$ in the following discussion,
but our main conclusions remain unaffected by this choice. In the
case of the Jain state we do not have a pseudopotential formulation
(a useful ansatz \cite{SS} that does not lead to a unique
zero-energy eigenstate is $\{0,V_{1},0,0,\ldots\}$).

\begin{figure}[htb]
\centering
\includegraphics[angle=270,scale=0.35]{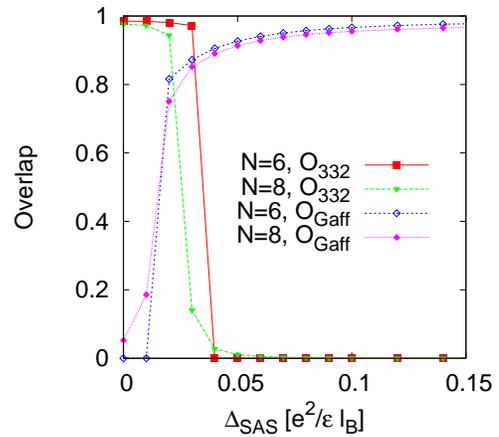}
\caption{(Color online) Overlaps between the exact Coulomb bilayer
ground state for $d=l_B$ and the 332 ($O_{332}$) and Gaffnian state
($O_{\rm Gaff}$), as a function of tunneling $\Delta_{SAS}$. Data
shown is for $N_e=6$ and $8$ electrons. Note that $O_{332}$ and
$O_{\rm Gaff}$ can not be directly compared due to the difference in
shift between the 332 state and the Gaffnian.} \label{f1}
\end{figure}
In Fig. \ref{f1} we present our results for the case of the bilayer Coulomb interaction
on the sphere with the bilayer distance $d$ equal to $l_{B}$. Overlaps of the exact state with the 332 state
and the Gaffnian are calculated as a function of tunneling $\Delta_{SAS}$. Separate diagonalizations
have been performed because the two trial states, 332 and Gaffnian, occur in slightly different
Hilbert spaces due to the mismatch in shift ($\delta=3$ and $\delta=4$, respectively).
Following the rapid destruction of the 332 state with the increase of $\Delta_{SAS}$,
the overlap with the Gaffnian state rises to the high value known from earlier studies in a single-layer model.\cite{SS, rbh}
This occurs at the point when the system is almost fully $x$-polarized.
Consequently, the overlap with the Jain state for large $\Delta_{SAS}$ is also high and
virtually indistinguishable from that of the Gaffnian on the scale of this figure.

\subsection{Torus}\label{sec_torus}

\begin{figure}[htb]
\centering
\includegraphics[angle=270,scale=0.35]{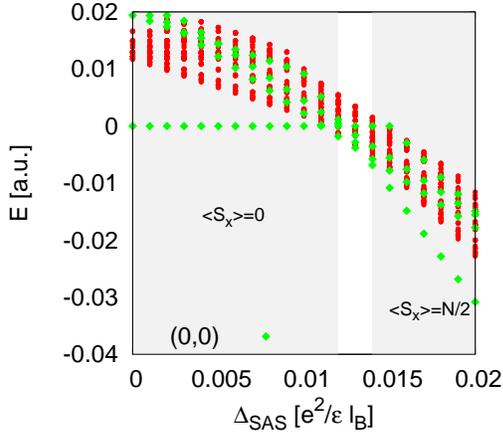}
\caption{(Color online) Energy spectrum of the SU(2)-symmetric 332 Hamiltonian on torus (in arbitrary units) for $N_e=8$ and aspect ratio 0.97.
The $\mathbf{k}=0$ levels that cross define regions of fully polarized $\langle S_x \rangle = N/2$ and unpolarized $\langle S_x \rangle = 0$ phases.} \label{f4}
\end{figure}

\begin{figure}[htb]
\centering
\includegraphics[angle=270,scale=0.35]{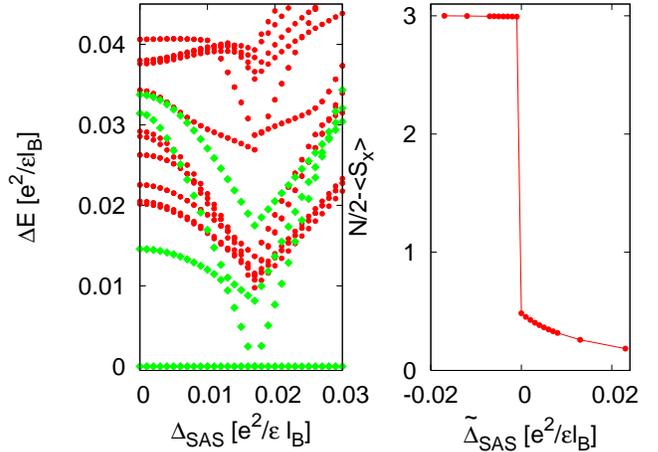}
\caption{(Color online) Energy spectrum relative to the ground state
of the Coulomb bilayer on torus, for $N_e=6$ electrons, $d=l_B$ and
aspect ratio 0.97 (left). We indicate the states characterized by
$\mathbf{k}=(0,0)$ Haldane pseudomomenta. Also shown is the
polarization $N/2 - \langle S_x \rangle$ as a function of tunneling
around the transition point $\Delta_{SAS}^C \approx
0.017e^2/\epsilon l_B$ (right).} \label{f0}
\end{figure}

\begin{figure}[htb]
\centering
\includegraphics[angle=270,scale=0.35]{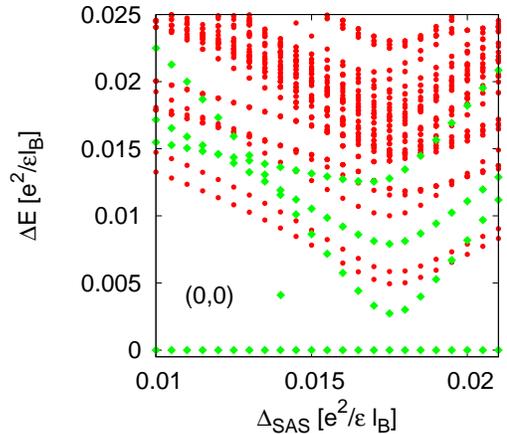}
\caption{(Color online) Energy spectrum relative to the ground state of the Coulomb bilayer on torus, for $N_e=8$ electrons, $d=l_B$ and aspect ratio 0.97. An approximate doublet of states with $\mathbf{k}=(0,0)$ Haldane pseudomomenta is formed around the transition point $\Delta_{SAS}\approx 0.018 e^2/\epsilon l_B$.} \label{f2}
\end{figure}

\begin{figure}[htb]
\centering
\includegraphics[angle=270, scale=0.35]{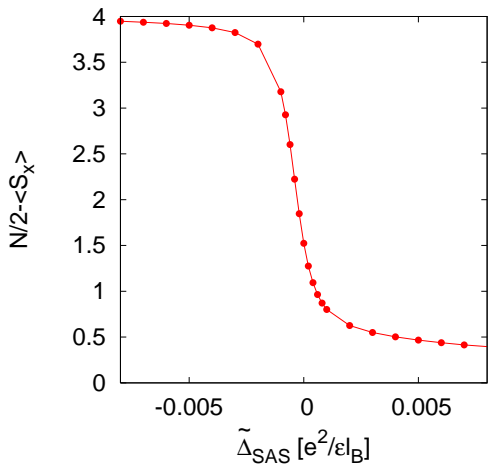}
\caption{(Color online) Polarization $N/2 - \langle S_x \rangle$ as a function of tunneling around the transition point for the system
of Fig. \ref{f2}.} \label{f3}
\end{figure}

\begin{figure}[htb]
\centering
\includegraphics[angle=270,scale=0.35]{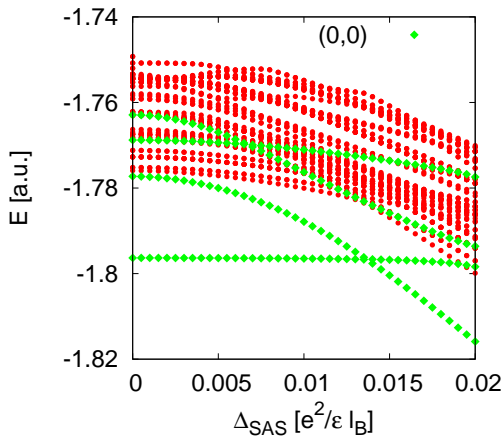}
\caption{(Color online) Energy spectrum of the Coulomb bilayer on torus (in arbitrary units) for $N_e=8$ and aspect ratio 0.5.} \label{f5}
\end{figure}

In the torus geometry, Figs. \ref{f4} -- \ref{f5}, trial states which describe
the same filling factor $\nu = p/q$ can be directly compared because the shift
iz zero. What is then characteristic of the Abelian states such as the 332 and
Jain's state, is that on the torus they only posses the
ground state degeneracy due to the motion of the center of mass of
the system, equal to $q$.\cite{haldanetorus} This is a trivial
degeneracy and we will mode out its presence in the data. In the
case of Gaffnian the degeneracy of the ground state is expected
\cite{SS,Eddy} to be doubled with respect to the trivial one i.e.
equal to $2 \times 5 = 10$. In the literature there is no consensus
that Gaffnian is a gapless state,\cite{SS, tokej} but if we can
establish that the nature of the lowest lying states is as expected
for the Gaffnian, we could nonetheless claim its presence at the
transition from the 332 to the Jain's state.

In Fig. \ref{f4} we plot the low energy spectrum of the 332
short-range Hamiltonian (Sec. \ref{sec_sphere}) on the torus for
$N=8$ electrons and close to the square unit cell (aspect ratio
$a/b=0.97$). We observe the 332 state, distinctly marked by its zero
energy, which remains unaffected by $\Delta_{SAS}$ until level
crossing is induced with the excited polarized state. We also
calculate the mean value of the $S_x$ projection of pseudo-spin
which plays the role of the ``order parameter" and has previously
been used to detect the transition between quantum Hall phases.
\cite{schliemann, nomura_yoshioka, gunnar} The state characterized
by $\langle S_x \rangle = N/2$ that becomes the ground state for
large tunneling develops into a Jain CF state, (\ref{jain}). This is
expected because the original system defined in terms of $V_c^{\rm
intra}(r), V_c^{\rm inter}(r)$ (\ref{intra}, \ref{inter}), in the
limit of very large tunneling becomes an effective one-component
model with the modified interaction $\left[ V_c^{\rm
intra}(r)+V_c^{\rm inter}(r)\right]/2$. \cite{zetal} For the
short-range 332 Hamiltonian, this is simply a $V_1$ pseudopotential
which yields a good approximation to Jain's state.\cite{SS}
Furthermore, as we vary the aspect ratio of the torus, we find the
following thin torus configuration $...0\; 1\; 0\; 0\; 1...$, which
is that of the Jain state. \cite{bergholtz}

Coulomb interaction shows stronger finite size effects that we exemplify with the spectra for $N=6$ (Fig. \ref{f0}) and $8$ electrons (Fig. \ref{f2}). In these calculations, we tune the aspect ratio to the same value of $a/b=0.97$
(slightly away from unity to avoid accidental geometric degeneracy) and distance between layers is set to $d=l_B$.
The incompressible states for small and large tunneling in Fig. \ref{f0} can be identified as the 332 and the Jain state,
with the transition between them occurring for $\Delta_{SAS}^C \approx 0.017e^2/\epsilon l_B$
when the levels cross (Fig.\ref{f0}, right), suggestive of the first order transition.
As a consequence, the polarization (``order parameter") $N/2-\langle S_x \rangle$ experiences a sharp discontinuity
at the point of transition (Fig. \ref{f0}, left). We stress that this level crossing occurs for a wide range of aspect
ratios of the torus and not only in the vicinity of the square unit cell.

On the other hand, for the larger system of $N_e = 8$ electrons
interacting with Coulomb interaction, we obtain the transition that
proceeds via level repulsion instead of level crossing, Fig.
\ref{f2}. We again  identify incompressible states for small and
large tunneling as the 332 and the Jain state, with the transition
between them occurring for $\Delta_{SAS}^C \approx 0.018e^2/\epsilon
l_B$. The states can be identified e.g. with respect to the Fig.
\ref{f4} by calculating overlaps. If we denote the ground state of
the short-range and Coulomb Hamiltonian for a given tunneling
$\Delta_{SAS}$ as $\Psi_{\rm short}(\Delta_{SAS})$ and $\Psi_{\rm
C}(\Delta_{SAS})$, respectively, we obtain the following overlap
$\langle \Psi_{\rm 332} | \Psi_C(\Delta_{SAS}) \rangle \equiv
\langle \Psi_{\rm short}(\Delta_{SAS}=0) |
\Psi_C(\Delta_{SAS}=0)\rangle \approx 0.95$. This means that the
Coulomb bilayer ground state is nearly the same as the 332 state,
assuming zero tunneling. Also, in the large tunneling limit, we
obtain e.g. $\langle \Psi_{\rm Jain} | \Psi_C (\Delta_{SAS}
\rightarrow \infty) \rangle \approx \langle \Psi_{\rm
short}(\Delta_{SAS} \rightarrow \infty) | \Psi_C(\Delta_{SAS}
\rightarrow \infty)\rangle \approx 0.948$, i.e. Jain's state.
The quantity which describes the density of the odd channel, $N/2 - \langle S_x \rangle$,
characterizes the transition by an approximately linear or even
step-like discontinuity as a function of $\tilde{\Delta}_{SAS} = \Delta_{SAS} -
\Delta_{SAS}^C$, Fig.\ref{f3}.
In the transition region, an approximate doublet of states with $\mathbf{k}=0$ Haldane pseudomomenta is formed (Fig. \ref{f2}).
Although the doublet has the expected quantum numbers of the Gaffnian\cite{Eddy}, the specific root configurations
in the thin torus limit\cite{bergholtz} cannot be unambiguously identified as those of the Gaffnian.
Both of the members of the doublet share the following thin torus configuration $...0\; 1\; 0\; 0\; 1...$, among other spurious patterns, which is that of the Jain state. Moreover, the member of the doublet higher in energy has a lower polarization $\langle S_x \rangle$ than the ground state. These facts suggest that the excited $\mathbf{k}=0$ state in the transition region is a spinful CF state rather than the
(polarized) Gaffnian.

For the long-range $N=8$ Coulomb system on the torus and the aspect ratio close to 1, the transition between the Jain and 332 state proceeds as an avoided level crossing or a smooth crossover without an obvious closing of the gap. The gap is expected to close in the thermodynamic
limit between the two distinct topological phases, although we are unable to perform a proper finite-size scaling of the gap due to the inaccessibility of the $N=10$ electron system.
However, for the short-range interaction that defines the 332 state as the zero-energy ground state and for the identical geometry of the torus ($a/b=0.97$), it appears that the gap indeed closes, Fig. \ref{f4}.
This difference between Figs. \ref{f4} and \ref{f2} can be attributed to the symmetry of the interaction. For the short range interaction used in Fig.\ref{f4}, $V_{1}^{\rm inter} = V_{1}^{\rm intra}$, hence it does not break the SU(2) symmetry. In this case, the tunneling part of the total Hamiltonian, being proportional to $S_{x}$ component, commutes with the interaction part and we expect level crossing which we indeed observe in Fig.\ref{f4}.  The interaction in the bilayer with $d=l_B$, on the other hand, breaks SU(2) invariance (Fig. \ref{f2}), but we can nevertheless show that the level crossing persists and can be induced by changing the aspect ratio of the torus away from unity. In Fig. \ref{f5} we show one such energy spectrum (without the ground state energy subtraction) when the aspect ratio is
equal to 0.5. The level crossing is induced by deforming the system towards the crystalline limit, when the Coulomb interaction is increasingly of short range. Note, however, that the states at $\Delta_{SAS}=0$ and $\Delta_{SAS}$-large are still 332 and Jain's, respectively (verified by the overlaps with the ground state of the short-range interaction and by their thin torus limit).

\section{Effective bosonic model}\label{sec_bosonic}

\subsection{Introduction}

High overlaps with the Gaffnian on the sphere around and after the
transition are a motivation for considering the system of
Chern-Simons (CS) transformed composite bosons\cite{zh, eziv}
$(\uparrow$ and $ \downarrow)$ that pair in the way of $p$-wave in a
picture of the underlying neutral sector physics. This bosonic
system is, by its very nature, unstable towards the ordinary Bose
condensation, as shown for the first time in Ref.
\onlinecite{gthesis}, and, as we will elaborate more, the pairing
may be realized only in its excited states or at a transition point.
As we will discuss in this Section, a simple underlying CS bosonic
picture of the 332 and Jain's state will be enlarged by $p$-wave
fluctuations. The fluctuations are expected to play a role near the
transition and in the description of the critical and excited
states, but not in the well-developed phases - the ground states
away from the transition. As we already pointed out in the preceding
section, the high Gaffnian overlaps are not to be taken as a proof
that we have the Gaffnian phase after the transition, in the
thermodynamic limit, but may serve as a motivation for discussing
the role for the Gaffnian as a critical state. More generally, as
the system is closer to the one-component limit, the theory may
inherit the pairing structure built in the Gaffnian state and this
is captured in the permanent state, (\ref{perm_state}). As we
mentioned in Section \ref{sec_system}, the connection between the
Gaffnian (\ref{gaff}) and the permanent state (a $p$-wave state of
bosons) (\ref{perm_state}) is the antisymmetrization. We assume that
the operation of antisymmetrization corresponds, in the language of
effective theory, to a tunneling term. \cite{zetal}

\subsection{Bosonic model}

To begin with, one may perform CS transformations in the
field-theoretical description of the system (\ref{def332}) that would leave, in the
mean field, $\uparrow$ and $\downarrow$ bosons that pair in the way
of a $p$-wave. At $\nu=2/5$, for no tunneling, in the presence of Coulomb
or suitable short range interaction, we expect that the bilayer
(two-component) system is described by 332 state. We know very well
how to define the CS transformation to bosons in these
circumstances, for the first time it was given in Ref.
\onlinecite{eziv}. It entails a transformation from electronic
$\Psi_{\sigma}$ fields (in Eq. (\ref{def332}) with $\Delta_{SAS}=0$)
to bosonic $\Phi_{\sigma}$ fields:
\begin{equation}
\Psi_{\sigma}(\mathbf{r}) = U_{\sigma}(\mathbf{r}) \Phi_{\sigma}(\mathbf{r})
\label{cs}
\end{equation}
where
\begin{equation}
U_{\sigma}(\mathbf{r}) = \exp\{-i\int\; d\mathbf{r}' {\rm arg}
(\mathbf{r} - \mathbf{r}') [  3 \rho_{\sigma}(\mathbf{r}') + 2
\rho_{-\sigma}(\mathbf{r}')]\}
\end{equation}
where ${\rm arg} (\mathbf{r} - \mathbf{r}')$ is the angle the vector
$\mathbf{r} - \mathbf{r}'$ forms with the $x$ axis. In the mean field
(when the fluctuations of gauge fields are
neglected) we, in fact, describe a system of $\uparrow$ and
$\downarrow$ bosons that interact. Therefore we have in the first
approximation two ordinary Bose condensates. By the virtue of the
Anderson-Higgs mechanism i.e. gauge fluctuations, the two Goldstone
modes become gapped and the two gapped bosonic systems describe
the two-component 332 system.

The complication comes when we consider the tunneling term as an extra
perturbation and an extra term in our starting Hamiltonian for
the electrons. The tunneling term is
\begin{equation}
H_{T} = - \lambda
\left(\Psi_{\uparrow}^{\dagger}(\mathbf{r})\Psi_{\downarrow}(\mathbf{r}) +
\Psi_{\downarrow}^{\dagger}(\mathbf{r})\Psi_{\uparrow}(\mathbf{r})\right),
\end{equation}
where $\lambda$ denotes the tunneling amplitude in this section.
Due to the CS transformation (\ref{cs}) this can not be
translated simply into the hopping of bosons because:
\begin{equation}
\Psi_{\sigma}^{\dagger} \Psi_{-\sigma} = \Phi_{\sigma}^{\dagger}
U_{\sigma}^{\dagger} U_{-\sigma} \Phi_{-\sigma}
\end{equation}
and only in the mean field approximation for which
\begin{equation}
U_{\sigma}^{\dagger} U_{-\sigma} \approx I \label{app}
\end{equation}
(where $I$ is the identity) we have a simple tunneling of bosons
i.e.
\begin{equation}
H_{T} \approx - \lambda\left(\Phi_{\uparrow}^{\dagger}(\mathbf{r})
\Phi_{\downarrow}(\mathbf{r}) + h.c.\right)
\end{equation}
The necessary assumption in Eq.(\ref{app}) is $\rho_{s} =
\rho_{\uparrow}(\mathbf{r}) - \rho_{\downarrow}(\mathbf{r})\approx 0$ i.e.
that the fluctuations in density in $\uparrow$ pseudospin parallel
the ones in $\downarrow$ or the fluctuations in the pseudospin
density are negligible.

Treating the residual interaction in a mean field manner i.e. taking
the Hartree-Fock and BCS decomposition, we come to the following form
of the Hamiltonian for the effective description of $\uparrow$ and
$\downarrow$ bosons around the $\mathbf{k} = 0$ point in the momentum space:
\begin{eqnarray}
H &=& \sum_{\mathbf{k}} \left( \sum_{\sigma} e \hat{b}^{\dagger}_{\mathbf{k} \sigma} \hat{b}_{\mathbf{k} \sigma}
- \lambda (\hat{b}^{\dagger}_{\mathbf{k} \uparrow} \hat{b}_{\mathbf{k} \downarrow}+ \hat{b}^{\dagger}_{\mathbf{k} \downarrow} \hat{b}_{\mathbf{k} \uparrow}) \right. \nonumber \\
 && \left. + d \; \hat{b}^{\dagger}_{\mathbf{k} \uparrow} \hat{b}^{\dagger}_{-\mathbf{k} \downarrow} + c
\; \hat{b}_{-\mathbf{k} \downarrow} \hat{b}_{\mathbf{k} \uparrow} \right) \label{ham}
\end{eqnarray}
where $ e = \epsilon_{\mathbf{k}} - \mu$ and $ d = c^{*}$ is the $p$-wave
order parameter function $ d \sim k_{x} - i k_{y}$.
The question of mutual statistics (between $\uparrow$ and
$\downarrow$ electrons and the ensuing composite bosons) may be raised but we will
assume that it is bosonic.

The Bogoliubov equations, $ [\alpha_{\mathbf{k}}, H] = E \alpha_{\mathbf{k}}$, where
\begin{equation}
\alpha_{\mathbf{k}} = u_{\uparrow} \hat{b}_{\mathbf{k} \uparrow} +
u_{\downarrow} \hat{b}_{\mathbf{k} \downarrow} + v_{\uparrow}
\hat{b}^{\dagger}_{-\mathbf{k} \uparrow} + v_{\downarrow}
\hat{b}^{\dagger}_{-\mathbf{k} \downarrow} \label{formsol}
\end{equation}
define the following matrix
\begin{displaymath}
\left[\begin{array}{cccc} e & - \lambda & 0 & -c \\
                          - \lambda & e & c  & 0 \\
                          0 & - d & - e & \lambda \\
                          d & 0 & \lambda & - e
\end{array} \right]
\end{displaymath}
for the eigenvalue problem. There are two pairs of eigenvalues:
\begin{eqnarray}
E + \lambda, E - \lambda,\; {\rm and}\; - E + \lambda, - E -
\lambda,
\end{eqnarray}
where $E = \sqrt{e^{2} - \Delta^{2}}, \Delta^{2} = d c$, with
the corresponding unnormalized eigenvectors:
\begin{eqnarray}
&& \alpha_\mathbf{k}^+ = \left[a, - a, 1, 1\right],\; \beta_\mathbf{k}^+=\left[a, a, - 1, 1\right],\; {\rm and} \label{eigenvecs+}\\
&&\alpha_\mathbf{k}^-=\left[b, - b, 1, 1\right],\; \beta_\mathbf{k}^-=\left[b, b, - 1, 1\right], \label{eigenvecs-}
\end{eqnarray}
where  $a = \frac{e + E}{d}$ and $b = \frac{e - E}{d}$.

\subsection{Bose condensate solution}\label{sec_bose}

The last two eigenvalues $-E\pm \lambda$ and vectors $\alpha_\mathbf{k}^-, \beta_\mathbf{k}^-$ (\ref{eigenvecs-})
define a pair of solutions in the form of (\ref{formsol}) and represent
well-defined excitations of the system.  The ground state can be expressed as
\begin{equation}
\exp\{\sum \frac{1}{b} \hat{b}^{\dagger}_{\mathbf{k}e} \hat{b}^{\dagger}_{-\mathbf{k}o}\}|0\rangle
\end{equation}
where $\hat{b}_{\mathbf{k}e} = \hat{b}_{\mathbf{k}\uparrow } + \hat{b}_{\mathbf{k}\downarrow }$ and $\hat{b}_{\mathbf{k}o} =
\hat{b}_{\mathbf{k}\uparrow } - \hat{b}_{\mathbf{k}\downarrow }$. We have for $\mu > 0$:
\begin{equation}
- E \pm \lambda \approx - \mu +
\epsilon_{\mathbf{k}} + \frac{\Delta^{2}}{2 \mu} \pm \lambda
\label{energy}
\end{equation}
i.e. ordinary non-interacting boson description where the tunneling
$\lambda$ defines the transition at $\lambda = \mu$ from the two
Bose condensates to a one Bose condensate (one disappears because
$\mu^{\rm eff} = \mu - \lambda < 0$ i.e. we have vacuum for these
particles). The mean field ground state in the $\mathbf{k} \rightarrow 0$ limit
is approximately constant ($ 1/b \sim d \rightarrow 0$) as it should be for the
effective description of the system with (two - becoming one) Bose
condensates.

This simple system in the presence of a short-ranged interaction, in
the channel that changes the sign of the effective chemical
potential, is described in Chapter 11.3 of Ref.\onlinecite{qpts}.
There $d_{\rm spatial} = 2$ was identified as an upper critical
dimension. Therefore we might expect in that case, with an
interaction, that the density of bosons in this channel for $
\lambda < \lambda_{c} = \mu$ vanishes linearly with $\mu - \lambda =
\mu^{\rm eff}$ as $\lambda \rightarrow \lambda_{c}$.

The quantum Hall system as a whole, together with the CS
fluctuations, may experience a  transition with the Bose condensates
becoming gapped via Anderson-Higgs mechanism(s) away from the
transition.

This analogy also motivates to consider that a viable composite
boson effective description of the $\nu = 2/5$ Jain's state is with
only one composite boson condensate and a Bose vacuum. This comes as
a natural consequence from our analysis and the multicomponent
approach to Jain's states. \cite{RGJ} From the results of the
experiments on the edge of FQH states,\cite{gray} it is justified to
assume an existence of a single charge mode that stems from a single
Bose condensate in an effective description.

\subsection{The role of the permanent}\label{sec_perm}

The transition may be discussed considering also the other pair of
eigenvalues from the eigenvalue problem:
\begin{equation}
E  \pm \lambda \label{secsol}
\end{equation}
and the excitations that they define (\ref{eigenvecs+}). It is
obvious from (\ref{energy}) that they are unstable and may describe
excited states. As particular solutions of the Bogoliubov equations
for the Hamiltonian defined in (\ref{ham}), the solutions described
by $E\pm \lambda$ and their corresponding eigenvectors
(\ref{eigenvecs+}) are non-unitary and non-physical because they are
related to the physical solutions (\ref{eigenvecs-}) by the
following non-unitary relationships:
$(\alpha_{-\mathbf{k}}^{-})^\dagger = i \beta_\mathbf{k}^{+}$ and
$(\beta_{-\mathbf{k}}^{-})^\dagger = - i \alpha_\mathbf{k}^{+}$
which imply:
$[(\beta_{\mathbf{k}}^{+})^\dagger,\beta_{\mathbf{k}}^{+}]=
[(\alpha_{\mathbf{k}}^{+})^\dagger,\alpha_{\mathbf{k}}^{+}]= 1$. We
assume a possibility that the description of the system is given by
$H$ and an additional term\cite{zetal} $\lambda N$ i.e. $ H +
\lambda N$, where $N$ is the number of particles. This term is of a
purely phenomenological origin; it is designed to regularize the
behavior at large $\lambda$ (compare with (\ref{excitations2})). It
can be incorporated in the previous description by a simple
redefinition of $e = \epsilon_{\mathbf{k}} - \mu$ into $e =
\epsilon_{\mathbf{k}} - \mu + \lambda$. This yields
\begin{eqnarray}
 E& = & \sqrt{(\epsilon_{\mathbf{k}} - \mu + \lambda)^{2} - \Delta^{2}}\nonumber \\
 & =&
 |\mu - \lambda| \sqrt{1 - \frac{(\mu - \lambda)}{(\mu -
 \lambda)^{2}}\; 2 \;\epsilon_{\mathbf{k}} - \frac{\Delta^{2}}{(\mu -
 \lambda)^{2}}},
\end{eqnarray}
and for large tunneling $\lambda > \mu$ we have
\begin{equation}
E \approx \lambda - \mu + \epsilon_{\mathbf{k}} -
\frac{1}{2}\;\frac{\Delta^{2}}{(\lambda - \mu)}.\label{excitations2}
 \end{equation}
The excitations (\ref{secsol}) become $E - \lambda \approx
\epsilon_{\mathbf{k}} - \mu$ and $ E + \lambda =
 \epsilon_{\mathbf{k}} + 2\; \lambda - \mu$, and we obtain Bose condensation in one channel
 and Bose vacuum in the other, as in the case of Sec. \ref{sec_bose}. Here,
 for $\lambda < \mu$ we allow the possibility that by this
 formalism we can describe an excited state of the system which is
 given by the Bogoliubov expression:
 \begin{equation}
\exp\{\sum \frac{1}{a} \hat{b}^{\dagger}_{\mathbf{k}e} \hat{b}^{\dagger}_{-\mathbf{k}o}\}|0\rangle
\label{gBog}
\end{equation}
and because $ \frac{1}{a} = \frac{e - E}{c}$, and $\sum
\frac{1}{a} \hat{b}^{\dagger}_{e \mathbf{k}} \hat{b}^{\dagger}_{o -\mathbf{k}} \sim \sum
\frac{1}{a} \hat{b}^{\dagger}_{\uparrow \mathbf{k}} \hat{b}^{\dagger}_{\downarrow -\mathbf{k}}$ we
have a $p$-wave paired permanent state in the long distance limit. This must be an excited
state because the excitations (\ref{secsol}) are unstable for $\lambda < \mu$
\begin{equation}
E \approx \mu -\lambda - \epsilon_{k} -
\frac{1}{2}\;\frac{\Delta^{2}}{(\mu -
 \lambda)}
 \end{equation}
At the transition $\lambda = \mu$ we have
\begin{equation}
E \approx \pm i \Delta_{0} |k| \label{prvajed}
\end{equation}
where $\Delta_{0}$ is defined by $\Delta^{2} = \Delta_{0}^{2}
k^{2}$. This defines a non-unitary system with complex values for
the excitations $E \pm \lambda$. If we neglect the presence of $\lambda$
for a moment, we can describe this system by a 2 + 1 dimensional
theory for bosons $\beta$ and $\gamma$ with the following
Hamiltonian:
\begin{equation}
H = \gamma \partial_{x} \beta .\label{drugajed}
\end{equation}
We quantize the system in the following manner:
\begin{eqnarray}
& \gamma = \sum ( \exp\{-i k x\} b_{k} + \exp\{i k x\}
a_{k}^{\dagger}) & \nonumber \\
& \beta = \sum ( - \exp\{i k x\} b_{k}^{\dagger} + \exp\{-i k x\}
a_{k}) & \label{trecajed}
\end{eqnarray}
and this reproduces the spectrum we have for $\lambda = 0$. This
system is closely related to the $\beta - \gamma$ ghost system in 1
+ 1 dimension or the CFT connected with permanent state \cite{rr}and
more generally Gaffnian \cite{SS} in its two-component formulation (\ref{gaff}).
The complete spectrum is reproduced by $ H = \gamma \partial_{x}
\beta + \lambda (\gamma \beta)$.

Therefore before reaching the strong tunneling limit and the
incompressible FQH state connected with the single BCS condensate in
this description at $\nu = 2/5$ (Jain's state), we may find a state
at the transition that evolves from an excited state. The excited
state above the 332 ground state (\ref{gBog}) is described in the
long-distance limit by a permanent times
the Abelian 332 factor, (\ref{perm_state}).  We note that the
permanent state carries the maximum pseudospin ($S = N_{e}/2,\;
\mathbf{S}^{2} = S (S + 1))$, because only such states (with also
$S_{z} = 0$) can be antisymmetrized completely in the coordinate
space and make a polarized electronic wave function just as in the
case of the permanent state and the ensuing Gaffnian wave function.
The 332 state, on the other hand, cannot be antisymmetrized
\cite{RGJ}, because it is a spin-singlet ($S = 0$). Depending on the
ground state evolution, the polarization of the system $(\langle S_x
\rangle)$ may either experience a jump across the transition or the
ground state may evolve smoothly into a $(\uparrow + \downarrow)$
polarized state. The state at the transition in the second case
might be Gaffnian - its description in the BCS formulation is that
of a state which evolves from the permanent under the effect of
tunneling which may mimic the antisymmetrization as in (\ref{gaff}).
But our analysis above (Eq. \ref{gBog} with the redefined $e$
and Eq. \ref{prvajed}) shows
that the system at the transition is still unpolarized and cannot
describe the Gaffnian.

\subsection{Discussion}
According to our numerical results in Figs. \ref{f0} - \ref{f3} in
the presence of Coulomb interaction a possible scenario is  the
scenario described in the Sec. \ref{sec_bose} with effectively one
of the two Bose condensates disappearing with the increase of
tunneling. If we include interactions in the simple bosonic model
they can smooth the transition (compare with results in Figs.
\ref{f2} and \ref{f3}). In Fig. \ref{f3}, we see the linear
dependence of the number of odd channel electrons on the tunneling
strength near the transition. In the $\mathbf{k} \rightarrow 0$
limit the density of the odd channel is equal to the density of the
vanishing Bose condensate. Therefore this linear dependence may stem
from the critical behavior of dilute bosons as described in
Ref.~\onlinecite{qpts}. $d_{\rm spatial} = 2$ is the upper critical
dimension in this case and we may expect a logarithmic correction to
the linear behavior as demonstrated in Ref.~\onlinecite{sss}, in the
case of a short range interaction among bosons. In our case Coulomb
interaction may be driving the fixed point for the short range
interactions into a mean field one with linear behavior. We
calculated the density-density correlator in the transition region,
but definite conclusion about the power of the decay of the
correlations could not be drawn because of the finite size effects.
A lower bound for the exponent that governs the decay with the
distance is equal to 2, as expected in the mean field.

Thus the bosonic model with interactions may lead to a second-order
transition with gradually disappearing bosons. In a more elaborate
description one may hope that Gaffnian will appear as a polarized
critical state before the polarized Jain state. But if the state at
the transition is partially polarized, as we find in exact
diagonalizations and effective bosonic model (without repulsive
interactions), even in the Coulomb case we may expect a first-order
transition or a smooth crossover without Gaffnian.

In the following we discuss implications of our analysis for the
effective bosonic description of the Jain's $\nu=2/5$ state. If, due
to tunneling, one Bose condensate indeed vanishes, the effective
description would then comprise only one Bose condensate and a Bose
vacuum.  On the other hand any effective description of quantum Hall
states must encompass the edge physics as the low energy physics of
these states happens on the edge. In the effective description based
on the usual picture with composite bosons \cite{Wen} of the $\nu =
2/5$ fractional quantum Hall edge, both charge and neutral edge
modes of two condensates propagate in the same direction as
relativistic particles and the discrepancy with respect to
experiments \cite{gray}, which detect only one (charge) mode, has to
be resolved \cite{WenLee}. In the effective description based on
composite fermions, \cite{LopFra} at $\nu = 2/5$ edge only the
charge mode is propagating, in agreement with the experiment, but
the reason why the neutral mode does not propagate is not obvious.
Here we suggest an effective picture of the neutral i.e.
multicomponent degrees of freedom of Jain's state at $\nu = 2/5$ via
a Bose vacuum. An edge excitation of the system that involves also
these, multicomponent, degrees of freedom, is accompanied by a
bosonic excitation of a vacuum that propagates, not
relativistically, but according to Schr\"odinger
equation\cite{qpts}, which in an effective description for certain
probes can be neglected with respect to the charge wave propagation
along the edge.


\section{Conclusions}\label{sec_conclusions}

We studied, by numerical and analytical means, the transition from
the two-component to a one-component quantum Hall state induced by
tunneling at the filling factor $\nu=2/5$. The transition is studied
in the presence of the Coulomb interactions appropriate for a
quantum Hall bilayer and a model short-ranged interaction
appropriate for the 332 Halperin's state. In exact diagonalizations
of small systems two possibilities for the transition are found: (a)
avoided level crossing, and (b) level crossing i.e. first-order
transition in the case of the Coulomb interaction and short range
interaction, respectively.

With respect to the appearance of the Gaffnian state in the
transition region between 332 and Jain state, we can conclude that
in finite systems this is only possible for the interaction that
breaks SU(2) invariance, like the Coulomb bilayer interaction. It is
an unlikely possibility, however, even for non-SU(2) invariant
interaction, because of the difficulty in establishing the thin
torus limit for the approximate $\mathbf{k}=0$ doublet found for the
torus with aspect ratio close to unity(Fig. \ref{f2}). In other
words, on the thin torus we observe only a ``half" \cite{SS} of the
Gaffnian physics that corresponds to Jain's state. So long as the
interaction is nearly SU(2) invariant, the transition occurs via
level crossing (Figs.\ref{f4}, \ref{f5}) and it is a first-order
transition between the unpolarized and polarized Abelian states.

Also, to probe the question of $p$-wave pairing and related Gaffnian
correlations at the transition we introduced an effective bosonic
model. We find that the transition in the presence of the Coulomb
interaction may be viewed as a transition from two Bose condensates
to a Bose condensate and a Bose vacuum. The outcome, with the Bose
vacuum, can serve as an effective description of the Jain state. In
the simple bosonic picture we find that the state at the transition
does not correspond to the polarized Gaffnian state, in accordance
with the exact diagonalizations.

\section*{Acknowledgments}
We thank Mark Goerbig and Nicolas Regnault for discussions. This
work was supported by the Serbian Ministry of Science under Grant
No. 141035.

\end{document}